\documentclass[12pt]{article}
\usepackage{amsmath}
\usepackage{amsthm}
\usepackage{amssymb}
\usepackage{amsfonts}
\usepackage{epic}
\usepackage{eepic}
 \usepackage{times}
\usepackage[matrix,arrow]{xy}
\usepackage{macros}
\usepackage{epsfig}
\usepackage{graphicx}
\usepackage{bm}

\usepackage{xcolor}

\usepackage[pagebackref=false]{hyperref}
\hypersetup{
    colorlinks = true,
    extension = notused,  
    linkcolor = blue,
    anchorcolor = red,
    citecolor = blue,
    filecolor = blue,
    pagecolor = red,
    urlcolor = blue
}

\theoremstyle{plain}

\theoremstyle{remark}

\newcommand{\sst}{\scriptscriptstyle}

\newcommand{\pa}{\partial}

\newcommand{\ra}{\to}

\newcommand{\fr}[2]{{\textstyle \frac{#1}{#2} }}

\newcommand{\al}{\alpha}

\newcommand{\be}{\beta}
\newcommand{\ga}{\gamma}

\newcommand{\de}{\delta}

\newcommand{\ep}{\epsilon}
\newcommand{\la}{\lambda}

\newcommand{\BBE}{S}
\newcommand{\bz}{{\bar{z}}}

\newcommand{\CA}{{\mathcal A}}

\newcommand{\CC}{{\mathcal C}}
\newcommand{\CD}{{\mathcal D}}

\newcommand{\CF}{{\mathcal F}}
\newcommand{\CG}{{\mathcal G}}

\newcommand{\CL}{{\mathcal L}}
\newcommand{\CM}{{\mathcal M}}
\newcommand{\CN}{{\mathcal N}}
\newcommand{\CO}{{\mathcal O}}
\newcommand{\CP}{{\mathcal P}}
\newcommand{\CQ}{{\mathcal Q}}

\newcommand{\CS}{{\mathcal S}}
\newcommand{\CT}{{\mathcal T}}

\newcommand{\CW}{{\mathcal W}}

\newcommand{\CZ}{{\mathcal Z}}

\newcommand{\SL}{{\mathsf L}}

\newcommand{\SO}{{\mathsf O}}

\newcommand{\BR}{{\mathbb R}}

\newcommand{\BC}{{\mathbb C}}

\newcommand{\BZ}{{\mathbb Z}}

\newcommand{\rf}[1]{(\ref{#1})}

\newcommand{\nc}{\newcommand}

\nc{\rnc}{\renewcommand} \nc{\beq}{\begin{equation}}
\nc{\eeq}{\end{equation}} \nc{\beqa}{\begin{eqnarray}}
\nc{\eeqa}{\end{eqnarray}}

\begin{document}
\title{Exact results on $\CN=2$ supersymmetric gauge theories}
\author{J. Teschner}
\address{
DESY Theory, Notkestr. 85, 22603 Hamburg, Germany}
\maketitle

The following is meant to give an overview over our special volume. The first three sections \ref{background}-\ref{use}
are intended to give a general overview over the physical motivations behind this direction of
research, and some of the developments that initiated this project. 
These sections are written for a broad audience of readers with interest in quantum field theory, 
assuming only very basic knowledge  of supersymmetric gauge theories and string theory. 
This will be followed in Section \ref{included} by a brief overview over the different chapters 
collected in this volume, while Section \ref{leftout} indicates
some related developments that we were unfortunately not able to cover here.

Due to the large number of relevant papers the author
felt forced to adopt a very restrictive citation policy. With the exception 
of very few original papers only review papers will be cited 
in Sections \ref{background} and \ref{news}. 
More references are given in later sections, but it still seems
impossible to list all papers on the subjects mentioned there.
The author apologises for any omission that results
from this policy. A citation of the form [V:$x$] refers to article number $x$ in this volume. 


\section{Background, history and context}\label{background}

\setcounter{equation}{0}

\subsection{Strong coupling behavior of gauge theories}

Gauge theories play a fundamental role in theoretical particle physics. They 
describe in particular the interactions that bind the quarks into hadrons. 
It is well understood how these interactions behave at high energies. This becomes
possible due to the phenomenon of asymptotic freedom: The effective strength of
the interactions depends on the energy scale, and goes to zero for large energies.
It is much less well understood how the interactions between quarks behave at low energies: The experimental evidence indicates
that the interactions become strong enough to prevent complete 
separation of the quarks bound in a hadron (confinement). The theoretical
understanding of this phenomenon has remained elusive.

When the interactions are weak one may approximate the resulting effects reasonably
well using perturbation theory, as can be developed systematically using the
existing Lagrangian formulations. However, the calculation of higher order effects
in perturbation theory gets cumbersome very quickly. It is furthermore well-known
that additional effects exist that can not be seen using perturbation theory. 
Exponentially suppressed contributions to the effective interactions are caused,
for example, by the existence of nontrivial solutions to the Euclidean equations
of motion called instantons. The task to understand the strong coupling behavior
of gauge theories looks rather hopeless from this point of view: It would require
having a complete resummation of all perturbative and non perturbative effects.
Understanding the strong coupling behaviour of general gauge theories remains 
an important challenge for quantum field theory. However, there exist 
examples in which substantial progress has recently been made on this problem:
Certain important physical quantities like expectation values of 
Wilson loop observables can even be calculated exactly.
What makes these examples more tractable is the existence of supersymmetry.
It describes relations between bosons and fermions which may imply that most 
quantum corrections from bosonic degrees of freedom cancel against similar
contributions coming from the fermions. Whatever remains may be exactly calculable.

Even if supersymmetry has been crucial for getting exact results up to now, 
it seems likely that some of the lessons that can be learned 
by analysing supersymmetric field theories will hold in much larger generality. 
One may in particular 
hope to deepen our insights into the origin of 
quantum field theoretical duality phenomena by analysing 
supersymmetric field theories,
as will be discussed in more detail below.
As another example let us mention that it was expected for 
a long time that instantons play a key role for the behaviour of 
gauge theories at strong 
coupling. This can now be illustrated beautifully 
with the help of the new exact results 
to be discussed in this volume. 
We believe that the study of supersymmetric field theories
offers a promising path to enter into the mostly 
unexplored world of non-perturbative phenomena
in quantum field theory.

\subsection{Electric-magnetic duality conjectures}

It is a hope going back to the early studies of gauge theories that there may 
exist asymptotic strong coupling 
regions in the gauge theory parameter space in which a conventional
(perturbative) description is recovered using a suitable new set of field variables. 
This phenomenon is called a duality.
Whenever this occurs, one may get access to highly nontrivial information about the 
gauge theory at strong coupling.

For future reference let us formulate 
a bit more precisely what it means to have a duality. Let us consider a family $\{\CF_z;z\in\CM\}$ of
quantum theories having a moduli space 
$\CM$ of parameters $z$. The quantum theory $\CF_z$ is 
for each fixed value of $z$
abstractly characterised by an algebra of observables 
$\CA_z$ and a linear functional on $\CA_z$ which assigns to  
each observable $\SO\in\CA_z$ its vacuum expectation value $\langle \SO\rangle_z$.
We say that $\{\CF_z;z\in\CM\}$ is a quantum field theory with 
fields $\Phi$ and action $S_\tau[\Phi]$ 
depending on certain parameters $\tau$  (like masses and
coupling constants) if there exists a point $z_0$ in the 
boundary of the moduli space $\CM$, a coordinate $\tau=\tau(z)$
in the vicinity of $z_0$, and a map $\CO$ 
assigning to each  $\SO\in\CA_z$ a functional $\CO_{\SO,\tau}^{}[\Phi]$
such that   
\begin{equation}\label{expz0}
\langle\,\SO\,\rangle_z\,\simeq \int [\CD\Phi]\;e^{-S_\tau[\Phi]}\,\CO_{\SO,\tau}^{}[\Phi]\,,
\end{equation}
where $\simeq$ means equality of asymptotic expansions around $z_0$ and the right hand side is defined 
in terms of the action $S[\Phi]$ using path integral methods.

We say that a theory with fields $\Phi$, action $S_\tau[\Phi]$ and parameters $\tau$
is dual to a theory characterised by similar data $S_{\tau'}'[\Phi']$ if there exists a family
of quantum theories $\{\CF_z;z\in\CM\}$ 
with moduli space $\CM$ having boundary points $z_0$ and $z_0'$ such that 
the vacuum expectation values of $\CF_z$ have an asymptotic expansion of the form
\rf{expz0} near 
$z_0$, and also an asymptotic expansion
\begin{equation}
\langle\,\SO\,\rangle_z\,\simeq \int [\CD\Phi']'\;e^{-S_{\tau'}'[\Phi']}\,\CO_{\SO,\tau'}'[\Phi']\,,
\end{equation}
near $z_0'$, with $\CO'$ being a map 
assigning to each  $\SO\in\CA_z$ a functional $\CO_{\SO,\tau'}'[\Phi']$.

A class of long-standing conjectures concerning the strong coupling behavior of gauge theories
are referred to as the electric-magnetic duality conjectures. Some of these
conjectures concern the infrared (IR) physics as described in terms of 
low-energy effective actions, others are about the full ultraviolet (UV)
descriptions of certain gauge theories. The main 
content of the first class of such
conjectures is most easily described for theories having an 
effective description at low 
energies involving in particular an abelian gauge field $A$ and some charged matter 
$q$.  
The effective action $S(A,q;\tau_{\rm\sst IR})$ will depend on an 
effective IR coupling constant $\tau_{\rm\sst IR}$.
The phenomenon of an electric magnetic duality would imply in particular
that the strong coupling behavior of such a gauge theory can be represented
using a dual action ${S}'({A}',{q}';{\tau}_{\rm\sst IR}')$ 
that depends on
the dual abelian gauge field ${A}'$ related to $A$ simply 
as
\begin{equation}
{F}_{\mu\nu}'\,=\,\frac{1}{2}\ep_{\mu\nu\rho\sigma}F^{\rho\sigma}\,. 
\end{equation}
The relation between the dual coupling constant 
${\tau}_{\rm\sst IR}'$ and ${\tau}_{\rm\sst IR}$ is also 
conjectured to be very simple,
\begin{equation}
{\tau}_{\rm\sst IR}'\,=\,-\frac{1}{\tau_{\rm\sst IR}}\,.
\end{equation}
The relation expressing ${q}'$ in terms of $q$ and $A$ may be very complicated,
in general. 
In many cases one expects that ${q}'$ 
is the field associated to solitons, 
localized particle-like excitations associated to 
classical solutions of the equations of motion
of $S(A,q;{\tau}_{\rm\sst IR})$. Such solitons are
usually very heavy at weak coupling but 
may become light at strong coupling where they may
be identified with fundamental particle 
excitations of the theory with action 
${S}'({A}',{q}';{\tau}_{\rm\sst IR}')$.

For certain theories there exist even deeper 
conjectures predicting dualities between
different perturbative descriptions of the full {\it ultraviolet} 
quantum field theories. Such conjectures, often referred to as
S-duality conjectures originated from the observations of 
Montonen and Olive \cite{MO,GNO}, and were subsequently refined in
\cite{WO,Os}, leading to the conjecture 
of a duality between the $N=4$ supersymmetric Yang-Mills theory
with gauge group $G$ and coupling $\tau$ one the one hand, and
the $N=4$ supersymmetric Yang-Mills theory
with gauge group ${}^{\rm L}G$ and coupling $-1/n_{\sst G}\tau$ on the 
other hand. ${}^{\rm L}G$ is the Langlands dual of a group $G$ having 
as Cartan matrix the transpose of the Cartan matrix of $G$, and $n_{\sst G}$ is the lacing number
\footnote{The lacing number $n_{\sst G}$ is equal to $1$ is the Lie-algebra
of $G$ is simply-laced, $2$ if it is of type $B_n$, $C_n$ and $F_4$, and $3$ if it is of type $G_2$.}
of the Lie algebra of $G$.

A given UV action $S$
can be used to define such expectation values perturbatively, as well as 
certain non-perturbative corrections like the instantons.
The question is whether all perturbative and non perturbative corrections
can be resummed to get the cross-over to the perturbation theory
defined using a different UV action $S'$. 

A non-trivial strong-coupling check for the S-duality conjecture in 
the $N=4$ supersymmetric Yang-Mills theory was performed in \cite{VW}.\footnote{The result
of \cite{Sen} furnishes a nontrivial check of a prediction following from the Montonen-Olive conjecture.}
Generalised S-duality conjectures have been formulated in \cite{G09} (see \cite{GaV} for a review) for a large class
of $\CN=2$ supersymmetric gauge theories  which are ultraviolet 
finite and therefore have well-defined bare UV coupling 
constants $\tau_r$.
It is of course a challenge to establish the validity of such
conjectures in any nontrivial example. 

\subsection{Seiberg-Witten theory}

A breakthrough was initiated by the discovery of exact results for the low energy
effective action of certain $\CN=2$ supersymmetric gauge theories by Seiberg and Witten \cite{SW1,SW2}. 
There are several good reviews on the subject, see
e.g. \cite{Bi,Le,P97,DPh,Tac} containing further references.\footnote{A fairly extensive list of references to the
early literature can be found e.g. in \cite{Le}.}

The constraints of $\CN=2$ supersymmetry restrict the low-energy physics 
considerably. As a typical example let us consider a gauge theory with
$SU(M)$ gauge symmetry. The gauge field sits in a 
multiplet of  $\CN=2$ supersymmetry
containing a scalar field $\phi$ in the adjoint representation of $SU(M)$.
$\CN=2$ supersymmetry allows parametric families of vacuum states. 
The vacuum states 
in the Coulomb branch can 
be parameterised by the vacuum expectation values of gauge-invariant functions
of the scalars like $u^{(k)}:=\langle{\rm Tr}(\phi^k)\rangle$, $k=2,\dots M$.
For generic values  of these quantities
one may describe the low-energy physics in terms of a Wilsonian effective action $S^{\rm eff}[A]$
which is a functional 
of $d=M-1$  vector multiplets $A_k$, $k=1,\dots,d$, having scalar components $a_k$ and 
gauge group $U(1)_k$, respectively.
The effective action $S^{\rm eff}[A]$ turns out to be completely determined by a single 
holomorphic function $\CF(a)$ of $d$ variables $a=(a_1,\dots,a_d)$ 
called the prepotential. It completely determines the
(Wilsonian) low energy effective action as $S^{\rm eff}=S^{\rm eff}_{\rm bos}+S^{\rm eff}_{\rm fer}$, where
\begin{equation}\label{Seffbos}
S^{\rm eff}_{\rm bos}\,=\,\frac{1}{4\pi} \int d^4x\;\Big( {\rm Im}(\tau^{kl})\pa_{\mu}\bar{a}_k\pa^{\mu}a_l
+\frac{1}{2}{\rm Im}(\tau^{kl})F_{k,\mu\nu}^{}F_{l}^{\mu\nu}+\frac{1}{2}{\rm Re}(\tau^{kl})F_{k,\mu\nu}^{}\tilde{F}_{l}^{\mu\nu}\Big)\,,
\end{equation}
while $S^{\rm eff}_{\rm fer}$ is the sum of all
terms containing fermionic fields, uniquely determined by
$\CN=2$ supersymmetry.  The $a$-dependent matrix $\tau^{kl}(a)$ in \rf{Seffbos} is the matrix of second derivatives of the
prepotential,
\begin{equation}
\tau^{kl}(a):=\pa_{a_k}\pa_{a_l}\CF(a)\,.
\end{equation}

Based on physically motivated assumptions about the strong coupling behavior 
of the gauge theories under consideration, Seiberg and Witten proposed a
precise mathematical definition of the relevant functions $\CF(a)$ for $M=2$. 
This type of description was subsequently generalised to large classes
of $\CN=2$ supersymmetric gauge theories including the cases with $M>2$.

The mathematics underlying the definition of $\CF(a)$ 
is called special geometry.
In many cases including the examples discussed above
one may  describe $\CF(a)$ using an auxilliary 
Riemann surface $\Sigma$ called the Seiberg-Witten curve which in suitable 
local 
coordinates can be described by a polynomial equation $P(x,y)=0$. 
The polynomial $P(x,y)$ has coefficients determined by 
the mass parameters, the gauge coupling constants, and 
the values $u^{(k)}$ parameterising the
vacua.
Associated to $\Sigma$ is
the 
canonical one form $\lambda_{\rm SW}=ydx$ on 
$\Sigma$. Picking a canonical basis for the first
homology $H_1(\Sigma,\BZ)$ of $\Sigma$, represented by curves $\al_1,\dots,\al_d$ and 
$\be_1,\dots, \be_d$
with intersection index $\al_r\circ \be_s =\de_{rs}$ one may consider
the periods
\begin{equation}
a_r\,=\,\int_{\al_r}\lambda_{\rm SW}\,,\qquad
a_r^{\rm D}\,=\,\int_{\be_r}\lambda_{\rm SW}\,.
\end{equation}
Both $a_r\equiv a_r(u)$ and $a_r^D\equiv a_r^D(u)$, $r=1,\dots,d$,  
represent sets of complex coordinates for the $d$-dimensional space of
vacua, in our example parameterised by $u=(u^{(2)},\dots,u^{(M)})$. 
It must therefore be possible to express $a^{\rm D}$ in terms of 
$a$. It turns out that the relation can be expressed using a function $\CF(a)$, $a=(a_1,\dots,a_d)$, 
from which the coordinates $a_r$ can be obtained via
$a_r^{\rm D}=\pa_{a_r}\CF(a)$. It follows that $\CF(a)$ is
up to an additive constant defined by $\Sigma$ and the choice of 
a basis for $H_1(\Sigma,\BZ)$. 

The choice of the field coordinates $a_k$ is not unique.
Changing the basis $\al_1,\dots,\al_d$ and 
$\be_1,\dots, \be_d$ to $\al_1',\dots,\al_d'$ and 
$\be_1',\dots, \be_d'$
will produce new coordinates 
${a}_r'$, ${a}'{}_r^{\rm D}$, 
$k=1,\dots,d$ along with a new
function ${\CF}'({a}')$ which is the prepotential 
determining a dual action ${S}_{\rm eff}'[{a}']$. The actions 
 ${S}_{\rm eff}[{a}]$ and  ${S}_{\rm eff}'[{a}']$ give us equivalent 
 descriptions of the low-energy physics. This gives an example for an  IR
 duality.

\subsection{Localization calculations of SUSY observables}\label{VEVs}

Having unbroken SUSY opens the possibility to compute some important quantities exactly using 
a method called localization \cite{W88}. This method forms the basis for much of the recent progress in this field. 

Given a supersymmetry generator $Q$ such that $Q^2=P$, 
where $P$ is the generator of a bosonic symmetry. Let  $S=S[\Phi]$ be an action 
such that $QS=0$. Let us furthermore introduce an auxiliary fermionic
functional $V=V[\Phi]$ that satisfies $PV=0$. We may then consider the path integral 
defined by deforming the action by the term $tQV$, with $t$ being a real parameter.
In many cases one can argue that expectation values of supersymmetric 
observables $\CO\equiv\CO[\Phi]$, $Q\CO=0$, defined by the deformed action 
are in fact independent of $t$, as the following formal calculation indicates. Let us consider
\begin{align}\notag
\frac{d}{dt}\int[\CD\Phi]\;e^{-S-tQV}\,\CO\,&=\,\int[\CD\Phi]\;e^{-S-tQV}QV\,\CO\\
&=\,\int[\CD\Phi]\;Q(e^{-S-tQV}\,V\,\CO)\,=\,0\,,
\label{localise}\end{align}
if the path-integral measure is SUSY-invariant, $\int[\CD\Phi]\;Q(\dots)=0$. This means that 
\begin{equation}
\langle \,\CO\,\rangle:=\int[\CD\Phi]\;e^{-S}\,\CO\,=\lim_{t\ra\infty}\int[\CD\Phi]\;e^{-S-tQV}\,\CO\,.
\end{equation}
If $V$ is such that $QV$ has positive semi-definite bosonic part, the only non-vanishing 
contributions are field configurations satisfying $QV=0$. There are cases where the space $\CM$ of solutions
of $QV=0$ is finite-dimensional.\footnote{In other cases $\CM$ may a union of infinitely many 
finite-dimensional components of increasing dimensions, as happens in the cases discussed in Section \ref{inst}.} 
The arguments above then imply that the expectation values can be 
expressed as an ordinary integral over the space $\CM$ which may be calculable.

The reader should note that this argument bypasses the actual definition of the path integral in
an interesting way. For the theories at hand, the definition of $\int[\CD\Phi]\;e^{-S-tQV}$ represents 
a rather challenging task which is not yet done. What the argument underlying the  localisation method
shows is the following: If there is ultimately {\it any} definition of the theory that ensures unbroken
supersymmetry in the sense that $\int[\CD\Phi]\,Q(\dots)=0$, the argument \rf{localise} will be applicable,
and may allow us to calculate {\it certain} expectation values exactly even if the precise definition 
of the full theory is unknown.

\subsection{Instanton calculus}\label{inst}

The work of Seiberg and Witten was based on certain assumptions on the strong coupling
behavior of the relevant gauge theories. It was therefore a major progress when
it was shown in \cite{N,NO,NY,BE} that the mathematical description for the 
prepotential conjectured by Seiberg and Witten can be obtained 
by an honest calculation of the quantum corrections to a certain 
two-parameter deformation of the prepotential to 
all orders in the instanton expansion.

To this aim it turned out to be very useful to define a regularisation
of certain IR divergences called Omega-deformation 
by adding terms to the action breaking Lorentz 
symmetry in such a way that a part of
the supersymmetry is preserved \cite{N}\footnote{The regularisation introduced in 
\cite{N} provides a physical interpretation of a regularisation 
for integrals over instanton moduli
spaces previously used in 
\cite{LNS,MNS1}.},
\begin{equation}
S \,\rightarrow S_{\ep_1\ep_2}=S+R_{\ep_1\ep_2}\,.
\end{equation}
One may then consider 
the partition function
$
\CZ
$
defined by means of the path integral 
defined by the action $S_{\ep_1\ep_2}$. As an example let us again consider 
a theory with $SU(M)$ gauge group.
This partition function $\CZ=\CZ(a,m,\tau;\ep_1,\ep_2)$ depends on the 
eigenvalues $a=(a_1,\dots,a_{M-1})$ of the
vector multiplet scalars at the infinity of $\BR^4$,
the collection $m$ of all mass parameters of the theory, 
and the complexified gauge coupling  $\tau$ formed out of
the 
gauge coupling constant $g$ and theta-angle $\theta$ 
as
\begin{equation}
\qquad\tau\,=\,
\frac{4\pi i}{g^2}+\frac{\theta}{2\pi}\,.
\end{equation}

The unbroken supersymmetry can be used to apply the localisation method
briefly described in Section \ref{VEVs}, here leading to 
the conclusion that the 
path integral defining $\CZ$ can be reduced to a sum of ordinary integrals
over instanton moduli spaces. The culmination of a long series of works\footnote{The results presented in 
\cite{N,NO} were based in particular on the previous work \cite{LNS,MNS1,MNS2}. 
Similar results were presented in
\cite{FPS,Ho1,Ho2,FP,BFMT}; for a review see \cite{TaV}.} were explicit formulae for the summands 
$\CZ^{(k)}(a,m;\ep_1,\ep_2)$ that appear in the resulting infinite series\footnote{The infinite series \rf{Zsum}
are probably convergent. This was verified explicitly for the example of
pure $SU(2)$ Super-Yang-Mills theory in \cite{ILTy}, and it is expected to follow
for UV finite gauge theories from the relations with conformal field theory to be
discussed in the next section.} of instanton corrections
\begin{equation}\label{Zsum}
\CZ(a,m,\tau;\ep_1,\ep_2)\,=\, \CZ^{\rm pert}(a,m,\tau;\ep_1,\ep_2)
\Bigg(1+\sum_{k=1}^{\infty}\,q^k\,
\CZ^{(k)}(a,m;\ep_1,\ep_2)
\Bigg)\,,
\end{equation}
with $q=e^{2\pi i\tau}$ in the ultraviolet finite cases,
while it is related to the running effective scale $\Lambda$ otherwise.
The explicitly known
prefactor $\CZ^{\rm pert}(a,m,\tau;\ep_1,\ep_2)$ is the product of the simple tree-level
contribution with a one-loop determinant. The latter is independent of the 
coupling constants $q_r$, and can be
expressed in terms of known special functions. 

In order to complete the derivation of the prepotentials proposed by 
Seiberg and Witten it then remained to argue that $\CF(a)\equiv 
\CF(a,m,\tau)$ is related to the 
partition function $\CZ$ as
\begin{equation}\label{FfromZ}
\CF(a,m;\tau) \,=\,-\lim_{\ep_1,\ep_2\ra 0} \ep_1\ep_2 
\CZ(a,m;\tau;\ep_1,\ep_2)\,,
\end{equation}
and to derive the mathematical definition of $\CF(a)$ proposed by 
Seiberg and Witten from the exact results on $\CZ(a,m;\tau;\ep_1,\ep_2)$ obtained in \cite{N,NO,NY,BE}.

\section{New exact results on $\CN=2$ supersymmetric field theories}\label{news}


\subsection{Localisation on curved backgrounds}\label{loccurved}

Another useful way to regularise IR-divergences is to consider the quantum field theory
on four-dimensional Euclidean space-times $M^4$ of finite volume. The finite-size effects encoded in the 
dependence of physical quantities with respect to the volume or other parameters of $M^4$ 
contain profound physical information. It has recently become possible to 
calculate some of the these quantities exactly.
One may, for example, consider gauge theories 
on a four-sphere $S^4$ \cite{Pe}, or more generally four-dimensional
ellipsoids \cite{HH},
\newcommand{\FE}{{{\BBE}^4}}
\begin{equation}
{\BBE}_{\ep_1,\ep_2}^4:=\,\{\,(x_0,\dots,x_4)\,|\,x_0^2+
\ep_1^2(x_1^2+x_2^2)+\ep_2^2(x_3^2+x_4^2)=1\,\}\,.
\end{equation}
The spaces ${\BBE}_{\ep_1,\ep_2}^4$ have sufficient symmetry for having an 
unbroken supersymmetry $Q$ such that $Q^2$ is the sum of a space-time symmetry plus 
possibly an internal symmetry.  Expectation values of supersymmetric observables
on ${\BBE}_{\ep_1,\ep_2}^4$ therefore represent candidates for 
quantities that may be
calculated by the localisation method.
Interesting physical quantities are the partition function on $\FE$, 
and the values of Wilson- and 't Hooft loop observables.
Wilson loop observables can be defined as path-ordered exponentials of the
general form
$W_{r,i}:={\rm Tr}\,\CP\exp\big[
\oint_{\CC}ds\;(i\dot{x}^\mu A_\mu^r+|\dot{x}|\phi^r)\big]$. 
The 't Hooft loop observables $T_{r,i}$, $i=1,2$, 
can be defined semiclassically 
by performing a path integral over field configurations 
with a specific singular behavior near a curve $\CC$ 
describing the effect of parallel transport of a magnetically
charged probe particle along $\CC$.
Choosing the support of the loop observables to be the 
circles $\CC_1$ or $\CC_2$ defined by
$x_0=x_3=x_4=0$ or $x_0=x_1=x_2=0$, respectively, one gets operators
commuting with part of the supersymmetries of the theory.

However, applying the localisation method to the field theories with $\CN=2$ 
supersymmetry is technically challenging  \cite{Pe,HH,GOP}.
A review of the necessary  technology and of some subsequent 
developments in this direction can be
found in the articles \cite{PV,OV}.
Appropriately modifying the action defining the theory under 
consideration on $\BR^4$
gives a $Q$-invariant action $S$ for the theory on  ${\BBE}_{\ep_1,\ep_2}^4$. 
A functional $V$ is found  in \cite{Pe} such that $QV$ is positive 
definite. The 
field configurations solving  $QV=0$ have constant values of the scalar fields, and vanishing values of all other 
fields. This means that the path integral reduces to an ordinary integral over scalar zero modes.
This phenomenon may be seen as a variant of the cancellations between contributions 
from fermionic and bosonic degrees of freedom that frequently occur in supersymmetric field theories,
leaving behind only contributions from states of zero energy. 

  
The results obtained by localisation \cite{Pe,GOP,HH} have the following structure:
\begin{itemize}
\item Partition functions:
\begin{equation}\label{Z}
{\mathbf Z}(m;\tau;\ep_1,\ep_2):=
\langle\,1\,\rangle_\FE^{S}\,=\,\int da \;|\CZ(a,m;\tau;\ep_1,\ep_2)|^2\,,
\end{equation}
where $\CZ(a,m;\tau;\ep_1,\ep_2)$ are the instanton partition functions briefly discussed in Section 
\ref{inst}. More details can be found in \cite{PV}.
\item
Wilson or 't Hooft loop expectation values:
\begin{equation}\label{loopVEV}
\langle\,\CL\,\rangle_\FE^{S}\,=\,\int da\;(\CZ(a,m;\tau;\ep_1,\ep_2))^*\;
\CD_{\CL}^{}\!\cdot\! \CZ(a,m;\tau;\ep_1,\ep_2)\,,
\end{equation}
where $\CZ(a,m;\tau;\ep_1,\ep_2)$ are the instanton partition functions described in Section \ref{inst}, and 
$\CD_{\CL}$ is a difference operator acting on the scalar zero mode variables collectively referred to by 
the notation $a$. The difference operators $\CD$ are pure multiplication operators
$\CD_{\CL}^{}=2\cosh(2\pi a/\ep_i)$ if $\CL$ is a  Wilson loop supported on $\CC_i$. These results are 
reviewed in \cite{OV}.
\end{itemize}
The integral over $a$ in \rf{Z}, \rf{loopVEV} is the integration over the scalar zero modes. One may interpret
these results as reduction to an effective quantum mechanics of these zero modes. From this point of view one
would interpret the instanton partition function $\CZ(a,m;\tau;\ep_1,\ep_2)$ as the wave-function $\Psi_\tau(a)$
of a state $|\tau\rangle_0$ in the zero-mode sub-sector defined by the path integral over field configurations
on the lower half-ellipsoid  ${\BBE}_{\ep_1,\ep_2}^{4,-}:=\{\,(x_0,\dots,x_4)\in{\BBE}_{\ep_1,\ep_2}^{4}\,;x_0<0\,\}$.
The expectation value \rf{loopVEV} can then be represented as 
\begin{equation}
\langle\,\CL\,\rangle_\FE^{S}\,=\,\langle\,\tau\,|\,{\CL}_0\,|\,\tau\,\rangle_0\,,
\end{equation}
where ${\CL}_0$ denotes the projection of the operator representing $\CL$ to the zero mode sub-sector.
This point of view is further 
discussed in \cite{TeV}.

Although the dynamics of the zero mode sub-sector is protected by supersymmetry, it 
captures very important non-perturbative information about the rest of the theory. Dualities
between different UV-descriptions of the gauge theory must be reflected in the zero mode dynamics, and can therefore 
be tested with the help of localisation calculations. But the 
definition of the full theory must
be compatible with these results, 
which is ultimately a consequence of unbroken supersymmetry. 
One may view the zero-mode dynamics as a kind of skeleton of the SUSY field theory. 
Whatever the QFT-``flesh'' may be, it must fit to the skeleton, and exhibit same dualities, for example.

The localisation method has furthermore recently been used to obtain 
exact results on 
some correlation functions in $\CN=2$ supersymmetric QCD \cite{BNP}.

\subsection{Relation to conformal field theory}\label{sec:AGT}

In \cite{AGT} is was observed that the results for partition functions of some four-dimensional supersymmetric
gauge theories that can be calculated by the method of \cite{Pe} are in fact proportional to something
known, namely the correlation functions of the  two-dimensional quantum field theory known 
as Liouville theory. Such correlation functions are formally defined by the 
path integral using the
action 
\begin{equation}
S^{\rm Liou}_b\,=\,\frac{1}{4\pi}\int d^2z\;\big[\,
(\pa_a\phi)^2+4\pi\mu  e^{2b\phi}\,\big]\,.
\end{equation}
Liouville theory has been extensively studied in the past motivated by the relations to two-dimensional quantum gravity 
and noncritical string theory discovered by Polyakov. It is known to be conformally invariant, as suggested by the 
early investigation \cite{CT}, and established by the construction given in \cite{Te}.
Conformal symmetry implies that the correlation functions can be
represented in a holomorphically factorized form. 
As a typical example let us consider 
\begin{equation}\label{holofact}
\big\langle \,e^{2\al_4\phi(\infty)}e^{2\al_3\phi(1)}
e^{2\al_2\phi(q)}e^{2\al_1\phi(0)}\,
\big\rangle^{\rm Liou}_b
=
\int _{\BR^+}\frac{dp}{2\pi} \; C_{21}(p)C_{43}(-p)
\big|
\CF_p\big[ 
{}^{\al_3}_{\al_4}
{}^{\al_2}_{\al_1}
\big]
(q)
\big|^2\,,
\end{equation}
where the 
conformal blocks
$\CF_p\big[ 
{}^{\al_3}_{\al_4}
{}^{\al_2}_{\al_1}
\big]
(q)
$
can be represented by power series of the form
\begin{equation}
\CF_p\big[ 
{}^{\al_3}_{\al_4}
{}^{\al_2}_{\al_1}
\big]
(q)\,=\,q^{\frac{Q^2}{4}+p^2-\al_1(Q-\al_1)-\al_2(Q-\al_2)}\bigg(1+\sum_{k=1}^{\infty}q^k\CF_p^{(k)}\big[ 
{}^{\al_3}_{\al_4}
{}^{\al_2}_{\al_1}
\big]\bigg)\,,
\end{equation}
having coefficients $\CF_p^{(k)}\big[ 
{}^{\al_3}_{\al_4}
{}^{\al_2}_{\al_1}
\big]$ completely defined by conformal symmetry  \cite{BPZ}.\footnote{A concise description of the definition of the conformal 
blocks can be found in \cite[Section 2.5]{TeV}.}
Explicit formulae for the coefficient functions
$C_{ij}(p)\equiv C(\al_i,\al_j,\frac{Q}{2}+ip)$, $Q=b+b^{-1}$,
have been conjectured in 
\cite{DO,ZZ}, and 
nontrivial checks for this conjecture were presented in \cite{ZZ}. 
A derivation of all these results follows from the free-field construction of Liouville theory
given in \cite{Te}.

In order to describe an example for the relations discovered in \cite{AGT}
let us temporarily restrict attention to the $\CN=2$ supersymmetric gauge theory 
often referred to as $N_f=4$-theory. This theory has field content consisting of an $SU(2)$-vector multiplet
coupled to four massive hypermultiplets in the fundamental representation of the gauge group.
The relation discovered in  \cite{AGT} can be written as
\begin{equation}\label{AGT-1}
\CZ(a,m;\tau;\ep_1,\ep_2)\propto
N_{21}(p)N_{43}(p)\CF_p\big[ 
{}^{\al_3}_{\al_4}
{}^{\al_2}_{\al_1}
\big]
(q)\,,
\end{equation}
where $|N_{ij}(p)|^2=C_{ij}(p)$. The factors of proportionality
dropped in \rf{AGT-1} are explicitly known, and turn out to 
be inessential. The parameters are identified, respectively, as 
\begin{subequations}\label{paramid}
\begin{align}
&b^2\,=\,\frac{\ep_1}{\ep_2}\,,\qquad
\hbar^2\,=\,\ep_1\ep_2\,,\qquad q=e^{2\pi i\tau}\,,\\
&p\,=\,\frac{a}{\hbar}\,,\qquad
\al_r\,=\,\frac{Q}{2}+i\frac{m_r}{\hbar},\,\qquad Q:=b+b^{-1}\,.
\end{align}
\end{subequations}
In order to prove \rf{AGT-1} one needs to show that 
the coefficients $\CZ^{(k)}(a,m;\ep_1,\ep_2)$ in \rf{Zsum} 
are equal to $\CF_p^{(k)}\big[ 
{}^{\al_3}_{\al_4}
{}^{\al_2}_{\al_1}
\big]$. This was done in \cite{AGT} up order $q^{11}$. 
A proof of this equality
for all values of $k$ is now available  \cite{AFLT}.

It furthermore follows easily from \rf{AGT-1} that the partition function
${\mathbf Z}(m;\tau;\ep_1,\ep_2)$ defined in \rf{Z} can be represented up to 
multiplication with an 
inessential, explicitly known function as
\begin{equation}\label{AGT-Z}
{\mathbf Z}(m;\tau;\ep_1,\ep_2)\,\propto\,
\big\langle \,e^{2\al_4\phi(\infty)}e^{2\al_3\phi(1)}
e^{2\al_2\phi(q)}e^{2\al_1\phi(0)}\,
\big\rangle^{\rm Liou}_b
\,.
\end{equation}

The relations between certain $\CN=2$ supersymmetric gauge theories and Liouville theory are most clearly
formulated in terms of the normalized expectation values of 
loop-observables 
\begin{equation}\label{normVEV}
\langle\!\langle\,\CL\,\rangle\!\rangle_\FE^{S}:=
\frac{\langle\,\CL\,\rangle_\FE^{S}}{\langle\,1\,\rangle_\FE^{S}}\,.
\end{equation}
To this aim let us note that the
counterparts of the loop observables within Liouville theory
will be certain nonlocal observables of the form
\begin{equation}
\SL_\ga:={\rm tr}\left[\CP\!\exp\left(\int_\ga \CA_y\right)\right]\,,
\end{equation}
where $\ga$ is a simple closed curve on $\BC\setminus\{0,q,1\}$, 
and $\CA$ is the flat connection
\begin{equation}
\CA:=\bigg(\begin{matrix} \,-\fr{b}{2}\pa_z\phi & 0 \\
\mu e^{b\phi} & \fr{b}{2}\pa_z\phi \end{matrix}\,\bigg)dz
+\bigg(\,\begin{matrix} \fr{b}{2}\pa_\bz\phi & \mu e^{b\phi} \\
0 & -\fr{b}{2}\pa_\bz\phi \end{matrix}\,\bigg)d\bz\,.
\end{equation}
Flatness of $\CA$ follows from the equation of motion.
Let us furthermore define normalized expectation values 
in Liouville theory schematically as
\begin{equation}
\langle\!\langle\,\CO\,\rangle\!\rangle^{\rm Liou}_b:=
\frac{\langle \;\CO\;e^{2\al_4\phi(\infty)}e^{2\al_3\phi(1)}
e^{2\al_2\phi(q)}e^{2\al_1\phi(0)}\,
\rangle^{\rm Liou}_b}
{\langle \,e^{2\al_4\phi(\infty)}e^{2\al_3\phi(1)}
e^{2\al_2\phi(q)}e^{2\al_1\phi(0)}\,
\rangle^{\rm Liou}_b}\,.
\end{equation}
We then have
\begin{equation}\label{loopAGT}
\langle\!\langle\,\CW\,\rangle\!\rangle_\FE^{S}\,=\,
\langle\!\langle\,\SL_{\ga_{s}^{}}\,\rangle\!\rangle^{\rm Liou}_b\,,\qquad
\langle\!\langle\,\CT\,\rangle\!\rangle_\FE^{S}\,=\,
\langle\!\langle\,\SL_{\ga_{t}^{}}\,\rangle\!\rangle^{\rm Liou}_b\,,
\end{equation}
where $\ga_{s}$ and $\ga_{t}$ are the simple closed curves encircling
the pairs of points $0,q$ and $1,q$ on $\BC\setminus\{0,q,1\}$,
respectively. A more detailed discussion can be found in
\cite{OV}
and \cite{TeV}.

\subsection{Relation to topological quantum field theory}\label{index}

The localisation method is also applicable in the case when the manifold $M^4$ has the form
$M^3\times S^1$ with supersymmetric boundary conditions for the fermions on the $S^1$.
In this case the partition function coincides with a quantity called index \cite{Ro,KMMR}, a trace 
${\rm tr}(-1)^{F}\prod_i\mu_i^{C_i}e^{-\be\{\CQ,\CQ^{\dagger}\}}$ 
over the Hilbert space of the theory on $M^3\times \BR$, with $F$ being the fermion number operator, $\CQ$ being 
one of the supersymmetry generators, and $C_i$ being operators commuting with $\CQ$. 
The index depends on parameters  $\mu_i$ called fugacities. It has originally been used to perform nontrivial checks 
of existing duality conjectures on field theories with $\CN=1$ supersymmetry \cite{DOs,SpV}. We will in the following
restrict attention to cases where the field theories have $\CN=2$ supersymmetry which are more closely 
related to the rest of the material discussed in this special volume.

As before one may use the 
localisation method to express the path integral for such manifolds as an integral
over the zero modes of certain fields,  with integrands obtainable by
simple one loop computations. This partition function can alternatively be computed by 
counting with signs and weights certain protected operators
in a given theory. If, for 
example, one takes $M^3=S^3$, the partition function of an ${\cal N}=2$ gauge theory with
gauge group $G$ and $N_f$ fundamental hypermultiplets takes the following form,
\begin{equation}\label{indexdef}
I({\bf b};p,q,t) = \oint [d{\bf a}]_G \, I_V({\bf a};p,q,t) \, \prod_{\ell=1}^{N_f} I_H({\bf a},{\bf b};p,q,t)\,, 
\end{equation}
where $[d{\bf a}]_G$ is the invariant Haar measure, we are using the notation $\{p,q,t,{\bf b}\}$ for the relevant
fugacities, and $I_V$ and $I_H$  are contributions coming from free vector multiplets and 
hypermultiplets, respectively.  The integral over ${\bf a}$ is roughly over the zero mode of the 
component of the gauge field in the $S^1$ direction. For more details see 
the article \cite{RR}. It is important to note that
the supersymmetric partition functions on $M^3\times S^1$ are independent 
of the coupling constants by an argument going back to \cite{W88}. Nevertheless, they are in general intricate 
functions of the fugacities and encode a lot of information about the protected spectrum of the theory.

There exists a relationship between the supersymmetric partition function on $M^3\times S^1$, the supersymmetric index, 
on  the one hand, and 
a topological field theory in two dimensions on the other hand~\cite{GPRR} which is somewhat analogous to
the relation of the $S^4$ partition function to Liouville theory discussed above. 
Let us consider the  example discussed above, ${\cal N}=2$ supersymmetric $SU(2)$ gauge theory with   $N_f=4$.  The supersymmetric index
of this theory can be represented in the form~\eqref{indexdef} noted above.
In the particular case when the fugacities are chosen to satisfy $t=q$,  
this index is equal~\cite{GRRY11} to a four point correlation function in a 
topological quantum field theory (TQFT) which can be regarded as a one-parameter deformation of
two-dimensional Yang-Mills theory with gauge grow $SU(2)$~\cite{AOSV},
\begin{equation}
I(b_1,b_2,b_3,b_4;p,q,t=q) =\prod_{\ell=1}^4 {\cal K}(b_\ell;q) \sum_{{\cal R}=0}^\infty C^2_{\cal R}  \prod_{\ell=1}^4 \chi_{\cal R}(b_\ell)\,.
\end{equation} Here $\chi_{\cal R}(x)$ is the character of a representation ${\cal R}$ of $SU(2)$.
The parameters $b_i$ are fugacities for the $\prod_{i=1}SU(2)_i$ maximal subgroup of the $SO(8)$ 
flavor symmetry group of the theory. This relation can be generalized to a large class of ${\cal N}=2$ theories 
and to indices depending on more general sets of fugacities~\cite{GRRY13,GRR}.

\section{What are the exact results good for ?}\label{use}

In the following we will briefly describe a few applications of the results
outlined above that have deepened our insights into supersymmetric field theories
considerably.

\subsection{Quantitative verification of electric-magnetic duality
conjectures}\label{duality}

The verification of the conjectures of Seiberg and Witten by the
works \cite{N,NO,NY,BE} leads in particular to a verification
of the electric-magnetic duality conjectures about the
low energy effective theories that were underlying the approach 
taken by Seiberg and Witten.\footnote{The IR duality conjectures 
can be used to describe the moduli space
of vacua as manifold covered by charts with local coordinates
$a_r$, $a_r^D$.  The transition functions between different charts define a Riemann-Hilbert problem. The solution
to this problem defines the function $\CF(a)$. It was shown in 
\cite{N,NO,NY,BE} that the series expansion
of $\CF(a)$ around one of the singular points on the moduli space of vacua satisfies 
\rf{FfromZ}. Taken together, one obtains a highly nontrivial check of the
IR-duality conjectures underlying Seiberg-Witten theory.}

Verification of UV duality relations like the Montonen-Olive duality
seems hopeless in general (see, however, \cite{VW}). 
However, in the cases where 
exact results on expectation values are available, as
briefly described in Subsection
\ref{sec:AGT},  one can do better.

In the case of the $N_f=4$ theory, for example,  one expects to find weakly 
coupled Lagrangian descriptions when the UV gauge coupling $q$
is near $0$, $1$ or infinity \cite{SW2}. Let us denote the actions
representing the expansions around these three values 
as ${S}_s$, ${S}_t$ and ${S}_u$, respectively. A particularly
important 
prediction of the S-duality conjectures is the exchange of the
roles of Wilson- and 't Hooft loops,
\begin{equation}
\label{S-duality2}
\langle\!\langle\, \CW \, \rangle\!\rangle_{{S}_s}^{}\,=\,
\langle\!\langle\, \CT \, \rangle\!\rangle_{{S}_t}^{}\,,
\qquad
\langle\!\langle\, \CT \, \rangle\!\rangle_{{S}_s}^{}\,=\,
\langle\!\langle\, \CW \, \rangle\!\rangle_{{S}_t}^{}\,.
\end{equation}
In order to check \rf{S-duality2} we may combine the results \rf{loopVEV} 
of the localisation computations
with the relation \rf{AGT-1} discovered in \cite{AGT}.
From the study of the Liouville theory one knows that the conformal blocks 
satisfy relations such as 
\begin{equation}\label{fusion}
\CF_p\big[ 
{}^{\al_3}_{\al_4}
{}^{\al_2}_{\al_1}
\big]
(q)\,=\,\int dp'\;F_{p,p'}\big[ 
{}^{\al_3}_{\al_4}
{}^{\al_2}_{\al_1}
\big]\,\CF_{p'}\big[ 
{}^{\al_1}_{\al_4}
{}^{\al_2}_{\al_3}
\big]
(1-q)\,,
\end{equation}
which had been established in \cite{Te}. These relations may now be re-interpreted as
describing  a resummation of the instanton expansion 
defined by action $S_s$ (the left hand side of \rf{fusion}) 
into an instanton expansion defined by the 
dual action $S_t$. This resummation gets represented
as an integral transformation with kernel $F_{p,p'}$. Using \rf{loopVEV}, \rf{AGT-1}, 
\rf{fusion}
and certain identities satisfied by the kernel $F_{p,p'}\big[ 
{}^{\al_3}_{\al_4}
{}^{\al_2}_{\al_1}\big]$  established in \cite{TV13}, one may now verify explicitly that the
S-duality relations \rf{S-duality2} are indeed satisfied.

In other words: Conformal field theory provides the techniques necessary to 
resum the instanton expansion defined  from a given action in terms of 
the instanton expansions defined from  a dual action.
At least for the class of theories at hand, these results  confirm in particular the long-standing 
expectations that the instantons play a crucial role for producing the cross-over between
weakly-coupled descriptions related by electric-magnetic dualities.

The electric-magnetic dualities can be also checked using the supersymmetric index. Although
the index does not depend on the coupling constants, in different duality frames it is given by 
different matrix integrals. Duality implies that these two matrix integrals evaluate to the same expression. In the relation of the index to TQFT discussed above, invariance under duality transformations in many cases follows from the associativity property of the TQFT.

\subsection{Precision tests of AdS-CFT duality}

Another famous set of duality conjectures concerns the behaviour of supersymmetric
gauge theories in the limit  where the rank of the gauge group(s) tends to infinity \cite{Ma}, see \cite{AGMOO} for a review. In this limit 
one expects to find a dual description in terms of the perturbative expansion of string theory 
on a background that is equal or asymptotic to five-dimensional Anti-de Sitter space. This duality predicts in some
cases representations for the leading strong-coupling behaviour of some gauge-theoretical observables
in terms of geometric quantities in supergravity theories. 

Some impressive quantitative checks of these duality conjectures are known in the case of 
maximal supersymmetry $\CN=4$ based on the (conjectured) integrability of the 
 $\CN=4$ supersymmetric Yang-Mills theory 
with infinite rank of the gauge group \cite{Bei}. 
Performing similar checks for theories with less supersymmetry 
is much harder. It is therefore worth noting 
that the localisation calculations of partition functions 
and Wilson loop expectation values described above have been used  
to carry out quantitative
checks of AdS-CFT type duality conjectures for some gauge 
theories with $\CN=2$ supersymmetry
\cite{BRZ,BEFP}.

It seems quite possible that the exact results described above 
can be used to carry out
many further precision tests of the AdS-CFT duality for 
$\CN=2$ supersymmetric field theories. The relevant backgrounds for string 
theory are not always known, but when they are known, one may use
the results obtained by localisation to check these generalised
AdS-CFT duality conjectures. Another result in this direction was 
recently reported in \cite{MP}.

The exact results can furthermore be used to study the phase 
structure of these gauge theories in the planar limit 
as function of the 't Hooft coupling. A surprisingly rich structure is found  in \cite{RZa,RZb}.
It seems that the full physical content of most of the data 
provided by the localisation calculations 
remains to be properly understood.

\subsection{Evidence for the existence of six-dimensional theories with (2,0)-supersymmetry}

Low-energy limits of string theory can often be identified with conventional 
quantum field theories. The string theorist's toolkit contains a large choice of
objects to play with, the most popular being compactifications and branes. 
One sometimes expects the existence of a low-energy limit with a certain amount 
of supersymmetry, but there is no known quantum field theory the limit could correspond to.
Such a line of reasoning has led to the prediction that there exists a very interesting class of interacting 
quantum field theories with six-dimensional $(2,0)$-superconformal invariance \cite{W95a,St,W95b}. These
theories have attracted a lot of attention over the last two decades, but not even the
field content, not to speak of a Lagrangian, are known for these hypothetical theories, see  \cite{Se,W09}
for reviews of what is known.

Nevertheless, the mere existence of such theories leads to highly non-trivial predictions,
many of which have been verified directly.  One could, for example, study the 
six-dimensional $(2,0)$-theories on manifolds of the form $M^4\times C$, where 
$C$ is a Riemann surface \cite{G09,GMN2} (some aspects are reviewed in  \cite{GaV,NV}). 
If $C$ has small area, one expects that the theory has
an effective description in terms of a quantum field theory on $M^4$. The 
resulting quantum field theory $\CG_C$ is expected to depend only on the 
choice of a hyperbolic metric on $C$ \cite{ABBR}, or equivalently (via the uniformisation theorem)
on the choice of a complex structure on $C$. The $N_f=4$-theory with four flavours 
mentioned above, for example, corresponds to $C=C_{0,4}$, which may be represented
as Riemann sphere with four marked points at $0,1,q,\infty$. One may use $q$ as 
parameter for the complex structure of $C_{0,4}$. When $q$ is near $0,1,\infty$, respectively,
it is natural to decompose $C_{0,4}$ into two pairs of pants by cutting along contours surrounding
the pairs of marked points $(0,q)$, $(q,1)$ and $(q,\infty)$, respectively. It turns out that
$q$ can be identified with the function $e^{2\pi i\tau}$ of the complexified gauge coupling 
$\tau=\frac{4\pi i}{g^2}+\frac{\theta}{2\pi}$ 
of the four-dimensional theory. The limits where $q$ approaches $0$, $1$ and $\infty$ are geometrically very
similar, but $q\ra 0$ corresponds to small gauge coupling, while $q\ra 1$ would correspond to a strong 
coupling limit. Note, on the other hand that 
the marked points at $0$ and $1$, for example, can  
be exchanged by a conformal mapping. 
This already suggests that there might exist a dual 
description having a complexified gauge coupling $\tau'$ such that $q'=e^{2\pi i\tau'}$ vanishes when
$q\ra 1$. The results described above provide a rather non-trivial quantitative check for this prediction.

Playing with the choice of $C$, and with the choice of the Lie algebra $\mathfrak g$ one can generate a large class
of interesting four-dimensional quantum field theories, and predict many non-trivial results about their physics \cite{G09,GMN2}.
The class of theories obtained in this way is often called class $\CS$.
Arguments of this type can be refined sufficiently 
to predict correspondences between four-dimensional gauge theories on $M^4$ and two-dimensional
conformal field theories on $C$ generalising the relations discovered in \cite{AGT}, see \cite{Y,CJ}. 
In the resulting generalisations of the relation
\rf{AGT-Z} one will find the correlation functions of the conformal Toda
theory associated to $\mathfrak g$ on the 
Riemann surface $C$, in general.
Considering the cases where 
$M^4=M^3\times S^1$, one may use similar
arguments to predict that the partition functions are related to correlation functions in a TQFT on $C$,
generalising the example
noted in Section \ref{index}. Such correlation function only depend on the topology of $C$,
corresponding to the fact that the partition functions on $M^4=M^3\times S^1$ are independent of 
exactly marginal coupling constants.
This will be discussed in more detail in \cite{RR}.

Other compactifications are also interesting, like $M^3\times C^3$ or $M^2\times C^4$, where $C^3$ and $C^4$
are compact three- and four-dimensional manifolds. Compactifying on $C^3$ or $C^4$ one gets
interesting quantum field theories on three- or two-dimensional manifolds $M^3$ or $M^2$, respectively. 
The origin from the six-dimensional $(2,0)$-theory may again be used to predict various nontrivial properties of the
resulting quantum field theories, including relations between three-dimensional field theories on $M^3$ 
and complex Chern-Simons theory on $C^3$ \cite{Y13,LY,CJ13}. Such relations are further discussed in \cite{DV}.

The six-dimensional $(2,0)$-theories are for $d=2,3,4$-dimensional 
quantum field theory therefore something like 
``Eierlegende Wollmilchs\"aue'', mythical beasts capable of supplying us with eggs, wool, milk and meat at the same time. 
The steadily 
growing number of highly nontrivial checks that the predictions following from its existence have passed
increase our confidence that such six-dimensional 
theories actually exist. Their 
existence supplies us with a vantage point from 
which we may get a better view on interesting parts of the landscape 
of supersymmetric
quantum field theories.

\subsection{Towards understanding non-Lagrangian theories}

There are many cases where strong-coupling limits of supersymmetric field theories
are expected to exist and to have a 
quantum field-theoretical nature, but no Lagrangian description of the resulting theories is known \cite{AD,APSW,AS}.  
The existence
of non-Lagrangian theories is an interesting phenomenon by itself. Certain non-Lagrangian quantum field
theories are expected to serve as elementary building blocks for the family of
quantum field theories obtained by compactifying $(2,0)$-theories \cite{G09,CD}.

The origin from a six-dimensional theory allows us to make quantitative predictions on some
physical quantities of such non-Lagrangian theories including the prepotential giving us the 
low-energy effective action, and the supersymmetric index giving us the protected spectrum 
of the theory. 

The results described in this special volume open the exciting perspective to go much further 
in the study of some non-Lagrangian theories. 
If the relation with two-dimensional conformal field theories continues to hold in the cases without known
Lagrangian descriptions, one may, for example compute the partition functions and 
certain finite-volume expectation values of loop operators in such theories. 
First steps in this direction were made in \cite{BMT,GT,KMST}.

\subsection{Interplay between (topological) string theory and gauge theory}

Superstring theory compactified on Calabi-Yau manifolds has two ``topological" relatives called the A- and the
B-model respectively. 
The ``topological" relatives are much simpler than the full superstring theories, but
they capture important information about the full theories like the coefficients of 
certain terms in the corresponding space-time effective
actions governing the low-energy physics. The A- and the B-model are not independent
but related by mirror symmetry.

The definition of the B-model topological string theory can be extended to so-called local Calabi-Yau $Y$,
defined (locally) by equations of the form
\begin{equation}
zw-P(u,v)\,=\,0\,,
\end{equation}
with $P(u,v)$ being a polynomial. Superstring theories on such local Calabi-Yau 
manifolds are expected
to have decoupling limits in which they are 
effectively represented by four-dimensional gauge 
theories\footnote{This limit is easier to define in the A-model, but the definition
can be translated to the B-model using mirror symmetry.}.
Describing four-dimensional gauge theory as decoupling limits of superstring theory is called
geometric engineering \cite{KLMVW,KKV,KMV}.
String-theoretic arguments \cite{N,LMN} predict that the instanton partition function (for $\ep_1+\ep_2=0$) 
coincides with the topological string partition function $\CZ^{\rm top}$ 
of the B-model on $Y$, schematically
\begin{equation}
\CZ^{\rm inst}\,=\,\lim_{\be\ra 0}\,\CZ^{\rm top}\,,
\end{equation} 
where $\be$ is related to one of the parameters for the complex structures on $Y$.
This prediction has been verified in various examples 
\cite{IK02,IK03,EK,HIV}. It
opens channels for the transport of information and insights from
topological string theory to gauge theory and back.
Interesting perspectives include:
\begin{itemize}
\item
Results from topological string theory may help to
understand 4d gauge theories even better, possibly
including non-Lagrangian ones. To give an example, let us note that the topological vertex \cite{AKMV,IKV}
gives powerful tools for the calculation of topological string partition functions. These results
give us predictions for the (yet undefined) instanton partition functions of non-Lagrangian theories,
and may thereby provide a starting point for future studies of the physics of such theories.
First steps in this direction were made in \cite{KPW,BMPTY,HKN,MP}.

\item Exact results on supersymmetric gauge theories may feed
back to topological string theory. As an 
example let us mention the development of the refined topological string,
a one-parameter deformation of the usual topological string 
theory that appears to exist for certain local Calabi-Yau manifolds, capturing 
nontrivial additional information. The proposal was initially
motivated by the observation that the instanton
partition functions can be defined for more general values of the 
parameters $\ep_1$, $\ep_2$ than the 
case $\ep_1+\ep_2=0$ corresponding to the usual topological string
via geometric engineering \cite{IKV,KW11,HK,HKK}.
There is growing evidence that such a deformation of the topological string  has a
world sheet realisation \cite{AFHNZa,AFHNZb}, 
and that the refinement fits well
into the conjectured web of 
topological string/gauge theory dualities \cite{AS12a,AS12b,CKK,NO14}. 
The relation with the holomorphic 
anomaly equation is reviewed in 
\cite{KWV}.

\item As another interesting direction that deserves further investigations
let us note that the  topological string partition functions
$\CZ^{\rm top}$ can be interpreted as  particular wave-functions in the quantum theory obtained by 
the quantisation of the moduli space of complex structures on Calabi-Yau manifolds,
as first pointed out in \cite{W93}, see \cite{ST} and references therein for 
further developments along these lines. By using the holomorphic anomaly 
equation one may construct $\CZ^{\rm top}$ as formal series in the topological string coupling constant $\la$,
identified with Planck's constant $\hbar$ in the quantisation of the moduli spaces of complex structures.
However, it is not known how to define  $\CZ^{\rm top}$ non-perturbatively in $\la$. 

On the other hand
it was pointed out above that the instanton partition functions are naturally interpreted as
wave-functions in some effective zero mode quantum mechanics to which the gauge theories in question
can be reduced by the localisation method. It seems likely that the effective zero mode quantum mechanics 
to which the gauge theories localise simply coincide with the quantum mechanics obtained from the quantisation
of the moduli spaces of complex structures which appear in the geometric engineering of the gauge theories
under considerations. These moduli spaces are closely related to  the moduli spaces of flat connections
on Riemann surfaces for the $A_1$ theories of class $\CS$.  The quantisation of these moduli spaces
is understood for some range of values of $\ep_1$, $\ep_2$ 
\cite{TeV}. Interpreting the results obtained thereby in terms of (refined) topological string theory
may give us important insights on how to construct $\CZ^{\rm top}$ non-perturbatively, at least for many local 
Calabi-Yau-manifolds.


\end{itemize}

\section{What is going to be discussed in this volume ?}\label{included}

Let us now give an overview of the material covered in this volume.

The first chapter  {{\it ``Families of $N=2$ field theories''}}  \cite{GaV}
by D. Gaiotto describes how large 
families of field theories with $\CN=2$ supersymmetry can be described  
by means of Lagrangian formulations,
or by compactification from the six-dimensional theory with $(2,0)$ 
supersymmetry on spaces of the form $M^4\times C$, with $C$ being 
a Riemann surface. The class of theories that can be obtained in this way
is called class $\CS$.
This description allows us to relate key aspects
of the four-dimensional physics of class $\CS$ theories to 
geometric structures on $C$.

The next chapter in our volume  is titled {\it ``Hitchin systems in $\CN=2$ field theory''} by A. Neitzke \cite{NV}.
The space of vacua of class $\CS$ theories on 
$\BR^3\times S^1$ can be identified as the 
moduli space of solutions to the self-duality equations
in two dimensions on Riemann surfaces studied by Hitchin. 
This fact plays a fundamental role for recent studies of the spectrum of 
BPS states in class $\CS$ theories, and it  
is related to the integrable structure underlying
Seiberg-Witten theory of theories of class $\CS$.
This article reviews important aspects of the role 
of the Hitchin system for the 
infrared physics of class $\CS$ theories.

In the following chapter {\it ``A review on instanton counting and W-algebras''} by Y. Tachikawa \cite{TaV},
it is explained how to compute  the instanton partition functions. The results can be written
as sums over bases for the equivariant cohomology of instanton moduli spaces. The known
results relating the symmetries of these spaces to the symmetries of conformal field theory 
are reviewed.

The Chapter 4 {{\it ``$\beta$-deformed matrix models and
the 2d/4d correspondence''}} by K. Maruyoshi \cite{MV} describes a very useful mathematical representation of the 
results of the localisation computations as integrals having a form familiar from the study of matrix models.
Techniques from the study of matrix models can be employed to extract important information
on the instanton partition functions
in various limits and special cases.

The Chapter 5 {\it ``Localization for $\CN=2$ Supersymmetric
Gauge Theories in Four Dimensions''} by V. Pestun \cite{PV} describes the 
techniques necessary to apply the localisation 
method to field theories on curved backgrounds like $S^4$, 
and how some of the results on partition functions outlined in 
Section \ref{loccurved} have been obtained. 

In our Chapter 6 {\it ``Line operators in supersymmetric gauge theories and the 2d-4d relation''}
by T. Okuda \cite{OV} it is discussed how to use localisation techniques 
for the calculation of expectation values of 
Wilson and 't Hooft line operators. The results establish direct connections between
supersymmetric line operators in gauge theories and the Verlinde line operators known from
conformal field theory. Similar results can  be used to strongly support 
connections to the quantum theories obtained from the  
quantisation of the  Hitchin moduli spaces.

Chapter  7 {\it ``Surface Operators''} by S. Gukov \cite{GuV} discusses a very 
interesting class of observables
localised on surfaces  that attracts steadily growing attention.
In the correspondence to conformal field theory some of these observables get 
related to a class of fields in two dimensions called
degenerate fields. These fields satisfy differential equations that
can be used to extract a lot of information on the correlation 
functions. Understanding the origin of these differential equations
within gauge theory may help explaining the 
AGT-correspondence itself.

There are further interesting quantities probing aspect of
the non-perturbative physics of theories of class $\CS$. The
Chapter 8
{\it ``The superconformal index of theories of class $\CS$''} by 
L. Rastelli, S. Razamat \cite{RR} reviews the 
superconformal index. It is often simpler to calculate than
instanton partition functions, but nevertheless allows one to perform 
many nontrivial checks of conjectured dualities. It turns out 
to admit a representation in terms of a new type of topological 
field theory associated to the Riemann surfaces $C$ parameterising 
the class $\CS$ theories.

The correspondence between four-dimensional supersymmetric gauge theories and 
two-dimensional conformal field theories discovered in \cite{AGT} has a very interesting relative, a correspondence 
between three-dimensional gauge theories and three-dimensional Chern-Simons theories
with complex gauge group.  It is related to the the correspondence of \cite{AGT}, but of interest in its own right. In order to see
relations
with the AGT-correspondences one
may consider four-dimensional field theories of class $\CS$ on half-spaces
separated by three-dimensional defects. The partition functions of the 
three-dimensional gauge theories on the defect turns out to be calculable by means of
localisation, and the results have a deep meaning
within conformal field theory or within the quantum theory of
Hitchin moduli spaces. 
How to apply the localisation method to (some of) the 
three-dimensional gauge theories that appear 
in this correspondence is explained in the Chapter 9
{\it ''A review on SUSY gauge theories on $S^3$''} by K. Hosomichi \cite{HV}.
The correspondences 
between three-dimensional gauge theories and 
three-dimensional Chern-Simons theory
with complex gauge group are the subject of Chapter 10
{\it ``3d Superconformal Theories from 
Three-Manifolds''}  by T. Dimofte \cite{DV}.

Chapter 11 {\it ''Supersymmetric gauge theories,
quantization of $\CM_{\rm flat}$, and Liouville theory''} by the author  \cite{TeV} describes 
an approach to understanding the AGT-correspondence by establishing a 
triangle of relations between the zero mode quantum mechanics obtained by localisation of 
class $\CS$ theories, the quantum theory obtained by 
quantisation of Hitchin moduli spaces, and conformal field theory. This triangle 
offers an explanation for the relations discovered in \cite{AGT}.

Some aspects of the string-theoretical origin of 
these results are discussed in the final Chapters of our volume.

Chapter 12 by {M. Aganagic and S. Shakirov, {\it ``Topological strings and 2d/4d 
correspondence''} \cite{AV}, describes one way to understand an important part of the 
AGT-correspondence in terms of a triality between four-dimensional gauge theory, 
the two-dimensional theory of its vortices, and conformal field theory. This triality is
related to, and inspired by known large $N$ dualities of the topological string.
It leads to a proof of some cases of the AGT-correspondence, and most 
importantly, of a generalisation of this correspondence to certain five-dimensional gauge theories.

In the final Chapter 13, {\it ``B-Model Approaches to 
Instanton Counting''},  D. Krefl, J. Walcher \cite{KWV} discuss the relation between the instanton partition 
functions and the partition function of the topological string from the perspective of the B-model. 
The instanton partition functions provide solutions to the holomorphic anomaly equations 
characterising the partition functions of the topological string. 

\section{What is missing ?}\label{leftout}

This collection of articles can only review a small part of the exciting recent progress on 
$N=2$ supersymmetric field theories.
Many important developments in this field could not be  covered here
even if they are related to the material discussed in our collection of articles in various ways.
In the following we want to indicate some of the  developments
that appear to have particularly close 
connections to the subjects discussed in this volume.

\subsection{BPS spectrum, moduli spaces of vacua and Hitchin systems}

BPS states are states in the Hilbert space of a supersymmetric field theory which 
are forming distinguished "small" representations of the supersymmetry algebra, a feature which
excludes various ways of "mixing" with generic states of the spectrum that would exist otherwise.
Gaiotto, Moore and Neitzke have initiated a vast program aimed at the study of the spectrum of 
BPS-states in the $\CN=2$ supersymmetric gauge theories $\CG_C$ of class $\CS$ \cite{GMN1,GMN2,GMN3}, 
see the article \cite{NV} for a review of some aspects.
To this aim it has turned
out to be useful to consider at intermediate steps of the analysis a compactification of the theories $\CG_C$ to 
space-times of the form $\BR^3\times S^1$. The moduli space of vacua of the compactified theory is 
"twice as large" compared to the one of $\CG_C$ on $\BR^4$, and it can be identified with Hitchin's 
moduli space of solutions to the self-duality equations on Riemann surfaces \cite{Hi}. 

The list of beautiful results that has been obtained along these lines includes:
\begin{itemize}
\item A new algorithm for computing the spectrum of BPS states which has a nontrivial, but piecewise constant
dependence on the point on the Coulomb-branch of the moduli space of vacua of $\CG_C$ on $\BR^4$.
The spectrum of BPS states may change along certain "walls" in the moduli space of vacua. Knowing the spectrum
on one side of the wall one may compute how it looks like on the other side using the so-called
wall-crossing formulae. Similar formulae were first proposed in the work of Kontsevich and Soibelman on 
Donaldson-Thomas invariants. 
\item Considering the gauge theory  $\CG_C$ compactified on $\BR^3\times S^1$ one may study natural line
operators including supersymmetric versions of the Wilson- or 't Hooft loop observables. Such observables can
be constructed using either the fields of the UV Lagrangian, or alternatively those of a Wilsonian IR effective action.
The vacuum 
expectation values of such line operators furnish coordinates on the moduli space $\CM$ of vacua of 
$\CG_C$ on $\BR^3\times S^1$ which turn out to coincide with natural sets of coordinates for  
Hitchin's moduli spaces.  The coordinates associated to observables defined in the IR reveal the structure
of Hitchin's moduli spaces as a cluster algebra, closely related to the phenomenon of wall-crossing in the
spectrum of BPS-states. Considering the observables constructed from the fields in the UV-Lagrangian one
gets coordinates describing the Hitchin moduli spaces as algebraic varieties. The relation between these
sets of coordinates is the renormalisation group (RG) flow, here protected by supersymmetry, and therefore
sometimes calculable \cite{GMN3}.
\end{itemize}
Even if the main focus of this direction of research is the spectrum of BPS-states, 
it turns out to deeply related to the relations discovered in \cite{AGT}, as is briefly
discussed in \cite{TeV}.

\subsection{Relations to integrable models}

It has been observed some time ago that the description of the prepotentials
characterising the low-energy physics of $\CN=2$ supersymmetric field theories provided by Seiberg-Witten theory
is closely connected 
to the mathematics of integrable systems \cite{GKMMM,MW,DW}. There are arguments indicating that
such relations to integrable models are generic consequences of $\CN=2$ supersymmetry: 
$\CN=2$ supersymmetry implies that the Coulomb branch of vacua carries a geometric structure 
called special geometry. Under certain integrality conditions related to the quantisation of 
electric and magnetic charges of BPS states one may canonically construct 
an integrable system in action-angle variables from the data characterising the special 
geometry of the Coulomb branch \cite{Fr}.

The connections between  four-dimensional field theories with  $\CN=2$-supersymmetry 
and integrable models 
have  been amplified enormously in a recent series of papers 
starting with \cite{NS}.\footnote{This paper is part of a program  initiated in \cite{NSa,NSb} 
investigating 
even more general connections between field theories with $\CN=2$-supersymmetry
and integrable models.}
It was observed that a {\it partial}
Omega-deformation 
of many $\CN=2$ field theories localised only on one of the 
half-planes spanning $\BR^4$ 
is related to the quantum integrable model obtained by quantising the classical integrable model related
to Seiberg-Witten theory.
The  Omega-deformation effectively 
localises the fluctuations to the origin of the half plane. This can be used to argue that
the low-energy physics can be effectively represented by a two-dimensional theory with 
$(2,2)$-supersymmetry \cite{NS} living on the half-plane in $\BR^4$ orthogonal 
to the support of the Omega-deformation. The supersymmetric vacua of the four-dimensional theory 
are determined by the 
twisted superpotential of the effective two-dimensional theory 
which can be calculated by 
taking the relevant
limit of the instanton partition functions. This was used  in \cite{NS} to argue that the supersymmetric
vacua are in one-to-one correspondence with the eigenstates of the quantum integrable model
obtained by quantising the 
integrable model corresponding to the
Seiberg-Witten theory of the four-dimensional 
gauge theory under consideration. 

This line of thought has not only lead to 
many new exact results on large families of four-dimensional $\CN=2$ gauge theories \cite{NRS,NP,NPS}, 
it has also created a new paradigm for the solution of algebraically integrable models.
More specifically
\begin{itemize}
\item For gauge theories $\CG_C$ of class $\CS$ an elegant description for the two-dimensional superpotential characterising
the low-energy physics in the presence of a partial Omega-deformation was given in \cite{NRS} in terms of the mathematics
of certain flat connections called opers
living on the Riemann $C$ specifying the gauge theory $\CG_C$.
\item In \cite{NP} the instanton calculus was generalised to a large class of $\CN=2$ gauge theories $\CG_\Gamma$
parameterised by certain diagrams $\Gamma$ called quivers. 
A generalisation of the techniques from \cite{NO}  allowed the authors to determine 
Seiberg-Witten type descriptions of the low-energy physics for all these theories, and to 
identify the integrable models whose solution theory allows one to calculate
the corresponding prepotentials.
\item The subsequent work \cite{NPS} generalised the results of \cite{NP} to the cases where
one has a one-parametric Omega-deformation preserving two-dimensional supersymmetry.
The results of \cite{NPS} imply a general correspondence between 
certain supersymmetric observables and the generating functions of conserved quantitates
in the models obtained by quantising the integrable models describing the 
generalisations of Seiberg-Witten-theory relevant for the gauge theories $\CG_\Gamma$.
\end{itemize}
The relations between these developments and the relations discovered in \cite{AGT} 
deserve further studies. One of the existing relations for $A_1$-theories of 
class $\CS$ is briefly discussed in the article \cite{TeV}. 
These results suggest that the AGT-correspondence and many related
developments can ultimately be
understood as consequences of the integrable 
structure in $\CN=2$ supersymmetric field theories. This point of view is 
also
supported by the relations between the quantisation of Hitchin moduli spaces 
and conformal field theory described in \cite{T10}.

\subsection{Other approaches to the AGT-correspondence} 

In this special volume we collect some of the basic results related to  the AGT-correspondence.
The family of results on this subject is rapidly growing, and many important developments
have occurred during the preparation of this volume. The approaches to proving or deriving the 
AGT-correspondences and some generalisations include
\begin{itemize}
\item Representation-theoretic proofs \cite{AFLT,FL,BBFLT} that W-algebra conformal blocks can be represented in 
terms of instanton partition functions. This boils down to proving existence of a basis
for W-algebra modules in which the matrix elements of chiral vertex operators
coincide with the so-called bifundamental contributions representing the main building blocks
of instanton partition functions.
\item Another approach \cite{MMS,MS} establishes relations between the series expansions for the 
instanton partition functions and the expressions provided by the free field
representation for the conformal blocks developed by Feigin and Fuchs, and Dotsenko and Fateev.
\item Mathematical proofs \cite{SV,MOk,BFN}
that the cohomology of instanton moduli spaces naturally carries a structure as 
a W-algebra module. This leads to a proof of the versions of the AGT-correspondence relevant for 
pure $\CN=2$ supersymmetric gauge theories for all gauge groups
of type $A$, $D$ or $E$.  The instanton partition functions get related to  
norms of Whittaker vectors in modules of W-algebras in these cases.
For a physical explanation of this fact see \cite{Tan}. Some aspects of this 
approach are 
described in \cite{TaV}.
\item Physical arguments leading to the conclusion that the six-dimensional $(2,0)$-theory
on certain compact four-manifolds or on four-manifolds $M^4$ with Omega-deformation
can effectively be represented in terms of two-dimensional 
conformal field theory \cite{Y,CJ}, or
as a $(2,2)$-supersymmetric sigma model with Hitchin target space \cite{NW}. 
\item Considerations of the geometric engineering of
supersymmetric gauge theories within string theory have led to the
suggestion that the instanton partition functions of the 
gauge theories from
class $\CS$ should be related to the partition functions of chiral free fermion
theories on suitable Riemann surfaces 
\cite{N}, see \cite{ADKMV,DHSV,DHS} for related 
developments\footnote{The relations between the topological vertex
and free fermion theories discussed in \cite{ADKMV} 
imply general relations between topological string partition 
functions of local Calabi-Yau manifolds, integrable models and
theories of free fermions on certain Riemann surfaces; possible
implications for four-dimensional gauge theories were
discussed 
in \cite{DHSV,DHS}.}. 
It was  proposed in \cite{CNO} that the relevant
theory of chiral free fermions is defined on the Riemann surface
$C$ specifying the gauge theories $\CG_C$ of class $\CS$.
These relations were called BPS-CFT correspondence in \cite{CNO}.
A mathematical link between BPS-CFT correspondence and the AGT-correspondence
was exhibited in \cite{ILT}.
\end{itemize}



\subsection{Less supersymmetry}

A very interesting direction of possible future research concerns possible generalisations of the results
discussed here to theories with less ($\CN=1$) supersymmetry. Recent progress in this direction
includes descriptions of the moduli spaces of vacua resembling the one provided by
Seiberg-Witten theory for field theories
with $\CN=2$ supersymmetry. 

The rapid growth of the number of publications on this 
direction of research makes it difficult to offer a 
representative yet concise 
list of references on this subject here.

\bigskip\noindent
{\bf Acknowledgements:} The author is grateful to D. Krefl, K. Maruyoshi, E. Pomoni, L. Rastelli,
S. Razamat, Y. Tachikawa and J. Walcher  for very useful comments and suggestions 
on a previous draft of this article.


\newcommand{\CMP}[3]{{ Commun. Math. Phys. }{\bf #1} (#2) #3}
\newcommand{\LMP}[3]{{ Lett. Math. Phys. }{\bf #1} (#2) #3}
\newcommand{\IMP}[3]{{ Int. J. Mod. Phys. }{\bf A#1} (#2) #3}
\newcommand{\NP}[3]{{ Nucl. Phys. }{\bf B#1} (#2) #3}
\newcommand{\PL}[3]{{ Phys. Lett. }{\bf B#1} (#2) #3}
\newcommand{\MPL}[3]{{ Mod. Phys. Lett. }{\bf A#1} (#2) #3}
\newcommand{\PRL}[3]{{ Phys. Rev. Lett. }{\bf #1} (#2) #3}
\newcommand{\AP}[3]{{ Ann. Phys. (N.Y.) }{\bf #1} (#2) #3}
\newcommand{\LMJ}[3]{{ Leningrad Math. J. }{\bf #1} (#2) #3}
\newcommand{\FAA}[3]{{ Funct. Anal. Appl. }{\bf #1} (#2) #3}
\newcommand{\TMP}[3]{{ Theor. Math. Phys. }{\bf #1} (#2) #3}
\newcommand{\PTPS}[3]{{ Progr. Theor. Phys. Suppl. }{\bf #1} (#2) #3}
\newcommand{\LMN}[3]{{ Lecture Notes in Mathematics }{\bf #1} (#2) #2}
\small  \setlength{\itemsep}{-3pt}

\paragraph{\large References to articles in this volume}
\renewcommand{\refname}{

\paragraph{Other references}
\renewcommand{\refname}{

\end{document}